\documentclass[10 pt, conference]{IEEEconf}
\usepackage[dvips]{graphicx}
\usepackage{cite,url,color,amsmath,graphicx,amssymb,times,subfigure,fancyhdr}

\input{epsf}

\newcommand {\beq} {\begin{equation}}
\newcommand {\eeq} {\end{equation}}
\newcommand {\barr} {\begin{array}}
\newcommand {\earr} {\end{array}}
\newcommand {\bear} {\begin{eqnarray}}
\newcommand {\eear} {\end{eqnarray}}
\newcommand {\bears} {\begin{eqnarray*}}
\newcommand {\eears} {\end{eqnarray*}}
\newcommand {\bn} {\begin{aligned}}
\newcommand {\en} {\end{aligned}}

\newcommand{\eat}[1]{}

\newtheorem{definition}{Definition}[section]
\newtheorem{lma}{Lemma}[section]
\newtheorem{prop}{Proposition}[section]

\def\for{{\textrm{ for }}}

\def\Prf{\vspace{2ex}\noindent{\bf Proof: }}
\def\endpf{\hfill$\diamond$}

\newcommand{\argmin}{\operatornamewithlimits{arg\,min}}

\begin{document}
\title{Unilateral Altruism in Network Routing Games with Atomic Players}
\author{Amar Prakash Azad \quad  John Musacchio\\Email: amarazad, johnm@soe.ucsc.edu
}
\maketitle

\begin{abstract}

We study a routing game in which one of the players unilaterally acts altruistically by taking into consideration the latency cost of other players as well as his own. By not playing selfishly, a player can not only improve the other players' equilibrium utility but also improve his own equilibrium utility. To quantify the effect, we define a metric called the Value of Unilateral Altruism (VoU) to be the ratio of the equilibrium utility of the altruistic user to the equilibrium utility he would have received in Nash equilibrium if he were selfish. We show by example that the VoU, in a game with nonlinear latency functions and atomic players, can be arbitrarily large. Since the Nash equilibrium social welfare of this example is arbitrarily far from social optimum, this example also has a Price of Anarchy (PoA) that is unbounded. The example is driven by there being a small number of players since the same example with non-atomic players yields a Nash equilibrium that is fully efficient.

\end{abstract}
\vspace{-1mm}
\section{Introduction}

The topic of network routing games has been studied extensively in recent years (e.g. \cite{orda,rou_bad,altman_Nash_distri}).
Some of these investigations consider network games for which there are so many users that the traffic of each user can be considered negligible -- the players for such games  are said to be non atomic. 
Other work considers routing games for which each player is the source of a significant amount of the network's traffic. These players are said to be atomic. 
The analysis of atomic routing games can be more complex than that of non atomic games since the routing choice of a single player can affect the delays seen by others.

The Price of Anarchy (PoA) \cite{deux} is one popular measure of the efficiency of equilibria for such games. 
It is defined as the ratio of  the maximum social welfare that could be achieved if a social planner selected the strategies of all players versus the lowest social welfare achieved in Nash equilibrium.
The literature shows that in many cases the PoA can be arbitrarily large, although there are several important classes of examples in which the PoA can be bounded. 
(For example, Roughgarden\cite{tim_anarchy_07}, shows that routing games with non-atomic users and affine latency functions have a PoA of no more than $4/3.$)

Naturally if players were not selfish, but instead all were striving to maximize social welfare, the equilibrium welfare could be much more efficient. 
However, an interesting question is what would happen if just one player chose not to be selfish and instead takes actions that consider the welfare of other players -- in other words if he behaves altruistically. 
In this work we investigate the effects of unilateral altruism. We find by way of example that the effect of unilateral altruism can be arbitrarily large.

%

A selfish user chooses a path (or a set of paths) that minimizes his total cost without considering that their selection may degrade the performance of the  other users in the system. 
However in practice players may not play like this.  Experimental studies \cite{Sobel,Levin:Altruism,ledyard:book} have shown that even in simple games in a controlled environment, players do not always act selfishly, for example they can show reciprocity, act out of spite, or be altruistic. Several explanations have been considered for such behaviour of players. Fehr \cite{Fehr,Fehr:book} argues that players act out of a perception of fairness and thus they consider a utility function that captures the costs of other players. 
Models of altruism and spite are discussed in  \cite{Levin:Altruism,ledyard:book,Chen:2008,Sharma:2009} etc. Most of the related models are discussed by Sobel in\cite{Sobel}. The fact that the social welfare observed in experimental trials is often better than that predicted by finding the Nash equilibrium of a model with selfish players is one reason to consider a model in which players act out of altruism. Since the players often interact over a longer period than that modeled in a one-shot game, players have a reason to play more cooperatively.

In this work, we focus on altruistic behaviour with atomic users. We consider a simple model where the perceived cost of a user is a linear combination of the other user's cost, as we proposed in\cite{altruism_wiopt}. We show that unilaterally altruism drastically increases the altruistic player's equilibrium utility. It is obvious that if all players are altruistic then the equilibrium utility can be improved, however it is surprising to observe a large improvement by a single player's altruistic action. 
 However, it needs to be emphasized that the improvement only happens if the other players change their strategies in accordance with the belief that the altruistic player will truly be altruistic. In some sense, the altruistic player has to ``credibly commit" to being altruistic.

In section \ref{s:smodel} we present the system model and key definitions.
We present an example to motivate the problem by showing very high price of anarchy in section \ref{s:me}.  Section \ref{s:2user} analyzes a two user selfish game. In section \ref{s:alt}, we introduce and analyze the altruistic routing game. In section \ref{s:est}, we extend our model to a more general topology with $n$ nodes and investigate the impact of altruism. 
Finally, we summarize with concluding remarks in section \ref{s:conclusion}. We skip  several details and proofs in the extended abstract due to space limitation, these details are made available at \cite{uni_hal}.

\vspace{-1mm}
\section{Basic Definitions and System Model}
\label{s:smodel}
We consider a network $\cal G=(\cal V,\cal L)$, where a set of nodes $\cal V$ are connected by set of directed links $\cal L \subseteq V\times V$. We denote $v_s\in{\cal S}\subseteq {\cal V}, v_d\in{\cal D}\subseteq  \cal V$  distinct nodes that are called source nodes and destination nodes, respectively. A route $R\in \cal R$ is a directed path with distinct nodes that connects a source and a destination. For every link $l\in \cal L$ and a route $R$, we write $l\in r$ whenever $l$ is a part of route $R$.

A set ${\cal I} = \{ 1, 2,..., I\}$ of users share the network $\cal
(V,L)$. We consider that each source node $v_s$ is associated with a user $i$, which ships a throughput demand of average data rate $r^i$ to a common destination associated with node $v_d$.  User $i$ splits its fixed demand
$r^i$ among the paths (routes) connecting the source to the destination, so
as to optimize some individual performance objective. We assume that there exist at least one route from $v_s$ to $v_d$.

Let $x_l^i$
denote the expected flow that user $i$ sends on link $l$. The vector of flows  $\mathbf{x}^i = {(x^i_l)}_{l\in{\cal L}}$ that user i sends onto each path is
called the routing strategy of user $i$. The set of strategies of user
$i$ that satisfy the user's demand and preserve its flow at all
nodes is called the strategy space of user $i$ and is denoted by
${\bf X}^i$, that is:
\vspace{-1mm}
 \bears {\bf X}^i =\{\mathbf{x}^i \in
\mathbb{R}^{|{\cal L}|} ; \sum_{l\in \textrm{Out}(v)} x^i_l =
\sum_{l\in \textrm{In}(v)} x^i_l+r^i_v,v\in {\cal V} \},
\eears
 where $r^i_s = r^i, r^i_d = -r^i$ and $r^i_v= 0$ for $v\neq v_s, v_d$ and $\textrm{Out}(v)$ and $\textrm{In}(v)$ the set of out-going and in-coming links, respectively. The
system flow configuration $\mathbf{x} = (x^1,...,x^I)$ is called a
\emph{routing strategy profile} and takes values in the product
strategy space ${\bf X} = \otimes_{i\in {\cal I}} {\bf X}^i$.

Every link $l\in \cal L$ is associated with a separable 
latency cost function $T_l(x_l):\cal R\rightarrow R$ which is assumed to be piecewise differentiable, convex, increasing, and $T_l(x_l)\geq 0$ for every $x_l >0$. The queueing network interpretation is that $T_l(x_l)$ is the delay incurred by traffic crossing link $l$, or alternatively it can be read  as the cost per unit of traffic when the load in link $l$ is $x_l$.

 \textbf{Latency function}: 
 We consider the following latency function types in this paper:
\begin{enumerate}
\item[$\bf{A:}$] Affine: $T_l(x_l)=a x_l+b$, where $a,b \geq 0$. 
\item[$\bf{E:}$] Elbow function: We define the function with offset $o$ in parametric form as $T_l(x_l)=\max\left(o, \left(\frac{L}{  \delta}\right)(x-r)+L\right)$. Unless stated otherwise, we use $o=0,0<\delta\ll1$, refer to Fig. \ref{fig:piecewise2cost}.
\end{enumerate}
The {$\bf E$} type latency function posses several features that makes in  mathematically tractable, e.g., piece-wise linearity, convexity. It turns out that a steep slope of elbow function causes users to react for even small flow changes which plays a key role in the latter sections. This is why, we designed the elbow function in a way so that the per-unit cost of traffic can increase very rapidly. This is also not unlike the delay in most queueing models such as $M/M/1$ delay function.  

\begin{figure}[tbp]
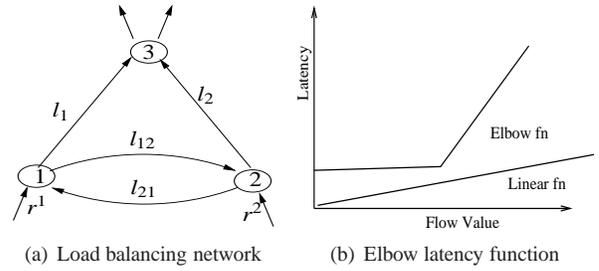

\begin{center}
\subfigure[
Load balancing network]{
 \resizebox{3.5cm}{3cm}{\input{2user.pstex_t}}
   \label{fig:lb} }
\subfigure[Elbow latency function]{
  \resizebox{4cm}{3cm}{\input{elbow.pstex_t}}
   \label{fig:elbow} }\vspace{-2mm}
\caption{Network topology and Latency function}
 \end{center}
 \vspace{-9mm}
\end{figure}

\textbf{Load Balancing (LB) Network:} We introduced a network topology called the ``load balancing network"\cite{altruism_wiopt}, where two users located one at each of the two nodes, called source nodes,  deliver their demand to a common destination node as pictured in Fig. \ref{fig:lb}. 

In this paper, we associate $\textbf{E}$ type latency with $l_1,l_2$ while we associate $\textbf{A}$ type latency with $l_{12},l_{21}$. One of the reasons of such modeling is that, typically source-destination links are standard queueing network links with limited capacity (approximated by $\textbf{E}$ type). The interconnecting/cross links are meant for balancing skewed loads on the local links. Thus, it is of high capacity but are expensive by provision (approximated by type $\textbf{A}$).   

{\textbf {User Cost}: }
Let $J_i({\bf x})$ be the cost of user $i$ under routing strategy
profile ${\bf x}$. The total cost to user $i$ is the sum of the separable link costs, given as
\vspace{-2mm}\bear
\label{e:Ji}
J_i({\bf x})= \sum_l x_l^i T_l(x_l), \,\, \textrm{for each } i.
\eear
The objective of each user $i$ is to find an admissible routing
strategy ${\bf x}^i\in {\bf X}^i$ so as to minimize his cost. Later we consider altruistic users, where the objective function takes into account of other user's cost also. 
\vspace{-2mm}
\subsection{Nash equilibrium}
Each user in this framework minimizes his own cost functions which
leads to the concept of Nash equilibrium. The minimization problem
here depends on the routing decision of other users, i.e., their
routing strategy \bears\label{nash_equilibrium}
\mathbf{x}^{-i}=(\mathbf{x}^{1},...,\mathbf{x}^{i-1},\mathbf{x}^{i+1},...\mathbf{x}^{I}),
\eears
\begin{definition}
A vector ${\mathaccent "7E {\bf x}}^i$, $i=1,2,...,I$ is called a
Nash equilibrium if for each user $i$, ${\mathaccent "7E {\bf x}}^i$
minimizes the cost function given that other users' routing
decisions are ${\mathaccent "7E {\bf x}}^{j}$, $j\not=i $. In other
words, \vspace{-2mm}
\bear\vspace{-2mm}
\label{eq:Nash_def} {J}_i({\mathaccent
"7E {\bf x}}^1, {\mathaccent "7E {\bf x}}^2,..., {\mathaccent "7E
{\bf x}}^I) =\min_{{\bf x}^i\in {\bf X}^i} {J}_i({\mathaccent "7E
{\bf x}}^1, {\mathaccent "7E {\bf x}}^2,..., {\bf x}^i,...,
{\mathaccent "7E {\bf x}}^I),&&\vspace{-2mm}\nonumber\\ i=1,2,...,I,&& \vspace{-2mm}
\eear 
where ${\bf X}^i$ is the routing strategy space of user $i$.
\end{definition}


 \textbf{Social Optimum:} A centralized or coordinated solution is referred to as social optimum\footnote{Existence of minima is due to compact strategy set and piece-wise continuous cost function. It is unique because of potential  function structure.}, which can be expressed as: \vspace{-2mm}
\bear \vspace{-2mm}
\mathbf{x}^{OPT}=\argmin_{\mathbf{x}\in{\mathbf{X}}} \sum_i J_i(\mathbf{x}).
\eear

\textbf{Price of Anarchy (PoA):} 
A well known way to quantify the equilibrium inefficiency of a non-cooperative game (in the sense of \cite{deux}). 
The Price of Anarchy is defined (\cite{deux,rou_bad}) as the ratio of worst case cost at Nash equilibrium when users are selfish to the cost at social optimum. Therefore,
\vspace{-2mm}\bear\vspace{-2mm}
 \label{eq:PoA}
PoA=\sup_{\textbf{x}\in \textbf{x}^{SE}} \frac{\sum_i J_i(\textbf{x})}{\sum_i J_i(\textbf{x}^{OPT})} \quad.
\eear
where $\textbf{x}^{SE}$ denotes the set of equilibrium flow configurations when users are  ``selfish".

\vspace{-1mm}
\section{Motivating Example}
\label{s:me}
In this section we illustrate the inefficiency of non-cooperative equilibrium with the help of a simple example and point out the key reason for the inefficiency to motivate the detailed investigation. We also note that the inefficiency appears only for the atomic users. 

Consider a LB network (in Fig. \ref{fig:lb}) with  two \textbf{selfish  users}, $u_1$ and $u_2$, at source nodes $1$ and $2$ respectively, having identical demand rates $r_1=r_2=r$, which is to be dispatched to the destination node $3$. 
Since, link $l_1$ is the most direct way for user 1 to reach the destination, we refer to link $l_1$ his ``local link". Similarly, we refer to link $l_2$ as user 2's local link. We refer to links $l_{12}$ and $l_{21}$ as ``cross'' links. 



\begin{figure}[tbp]
\begin{center}
\resizebox{3.5cm}{3cm}{\input{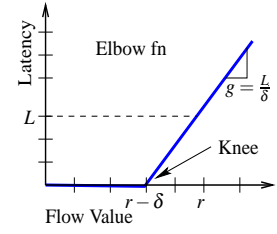}}
\vspace{-3mm}
\caption{Link latency function:  $L_2(x)=\frac{L}{  \delta}(x-r)+L,\,\for\, r-  \delta\le x $ and the parameters are: $  r=1,L=0.1,   \delta=10^{-3},g=\frac{L}{ \delta}=100$.  }
\label{fig:piecewise2cost}
\end{center}
\vspace{-10mm}
\end{figure}

\textbf{ Example}: 
Let the latency function be as shown in Fig. \ref{fig:piecewise2cost}, which is  expressed as:
\vspace{-2mm}
\bear
T_{l}(x)=
\begin{cases}
\max(0,L_2(x)), & \textrm{ for } l=l_1,l_2\\
 c, &\textrm{ for } l=l_{12},l_{21}.
\end{cases}\vspace{-2mm}
\eear
We can easily observe that  the social optimum is achieved when users use only local links. 
This is so because any positive flow on cross links add additional latency while retaining the total latency (due to flow) on links $l_1$ and $l_2$ remains unchanged. However, when users are selfish, each user optimize its own utility.  We illustrate by way of iteratively finding the best responses of the two selfish players why they end up putting a large fraction of their traffic on the cross links, which is the cause of inefficiency. 

 Let the initial flow be $\{  r,  r\}$,  and the corresponding initial cost of users are $\{L,L\}=\{0.1,0.1\}$.
  Let $u_1$ pushes $\delta$ amount of flow to $l_2$ to bring his local link flow below the knee (see Fig. \ref{fig:piecewise2cost} ), which results in $T_1(.)=0$.  On the other hand, the total flow in $l_2$ raised to $r+\delta$ which results in $T_2(.)=L+\frac{L}{\delta}\delta=2L$. The updated cost of $u_1$ is $J_1=\delta c+ \delta \cdot 2L=0.12$, which is much lower than previous cost $0.1$. However, the cost of $u_2$ has increased from $0.1$ to $(1-10^{-3})\approx 0.2$, almost doubled. In turn $u_2$ will reciprocate by pushing $2 \delta$ unit of flow to $l_1$ to bring his local link latency below the knee, which reduces his cost drastically to $2\delta\cdot (2L+c) \approx 0.003$, but raises the cost of $u_1$ to $(1-\delta)2L+\delta c \approx 0.2$. Due to this raise in cost, again $u_1$ will push $2\delta$ amount of flow to $l_2$ in order to bring his local link latency below the knee. 
  This process will keep repeating till the flow configuration  reaches to $x^{SE}\approx\{0.5, 0.5 \}$,  after which none of the users benefit by deviating even a small amount of traffic from their local link. In other words, $x^{SE}$ is the Nash equilibrium. The total cost incurred by any user at $x^{SE}$ is $\approx0.5\cdot(0.1+1)=0.55$, which is much higher ($\approx 50$ times) than the cost at social optimum ($=0.1$), 
  i.e.,  $PoA \approx 50$. 

We demonstrated by a simple LB Network example that selfish routing with atomic users can be significantly inefficient. In the following, we show that it can be even worse and can be arbitrarily large for atomic users. We also show that this is not the case with non-atomic users, suggesting that it is the ability of a user to chase the latency of the links with his strategy that drives the example. 

\vspace{-1mm}
\section{Two User Symmetric Game}
\label{s:2user}
In this section we establish results showing the inefficiency of selfish routing games by analyzing the two user LB network game. In the later sections, we construct and analyze the model in which users are altruistic, which exposes our key findings.  

We consider a sequence of LB network examples indexed by $m\geq 1$. 
Consider a LB Network of three nodes as shown in Fig. \ref{fig:lb} with the $\textbf{E}$ type and $\textbf{A}$ type latency functions as follows:
\vspace{-2mm}
\bear \label{e:Tl}
\hspace{-2mm}T_{l}(x)=
\begin{cases}
\max\left(0, \left(\frac{L}{  \delta_m}\right)(x-  r)+L\right), &\hspace{-2mm}\textrm{for } l_1,l_2,\\
 c_m, &\hspace{-2mm}\textrm{for } l_{12},l_{21}.
\end{cases}
\eear
where  $c_m>L> \delta_m>0$ and $c_m<r\frac{L}{ \delta_m}$ for all $m$ . The sequence ${\delta_m}$ decreases and tends to zero such that the gradient $\frac{L}{\delta_m} \rightarrow \infty$ as $m \rightarrow \infty$. Similarly, $c_m$ is an increasing sequence such that latency of cross link can be made very high for sufficiently large $m$. One example of such sequence can be $\delta_m=\delta^m$ and $c_m=c^m$. It turns out to be important that the gradient $\frac{L}{\delta_m}$ be larger than the cross-link latency $c_m$, but the cross-link latency $c_m$ be bigger than the latency of the local links when the flow is $r$, which is $L$. 

Using \eqref{e:Ji}, the cost of users are given by,
\vspace{-2mm}
\bear \label{eq:J1}
&&\hspace{-10mm} J_i(\mathbf{x})=x_i^i T_i(x_i) +\bar x_i^i(T_{ij}(\bar x^i_i)+T_j(x_j)),\,\,\for \,i,j=1,2.
\eear
where, $\bar x_i^i=r^i-x_i^i$ 
We often denote $T_{l_i}(.)$ by $T_i(.)$ for the ease of notation.
\vspace{-1mm}
  \subsection{Selfish Routing Game with Atomic Users }
  \begin{lma} 
  \label{l:1} 
  Consider two atomic users of identical demands on a load balancing network (Fig. \ref{fig:lb}) at node $1$ and node $2$ with the associated latency given by \eqref{e:Tl}.
  In the two user selfish routing game, there exists a unique Nash equilibrium, given by     
\vspace{-2mm}
  \bear
  \label{eq:X11*}
{ \textbf{x}^{SE}}&=&\{x_1^{1^*}, x_2^{2^*}\}=\left\{  r /2+\zeta_m, r /2+\zeta_m\right\},\\  
 J_i^{SE}&=& r L+( r/2-\zeta_m)c_m, \,\, \textrm{for } i=1,2.
\eear
where $\zeta_m=\frac{1}{2}(\frac{c_m}{L/  \delta_m})$ is 
the ratio of  unit latency of cross links to the marginal latency of local links. By abuse of notation, above we denoted $\textbf{x}^{SE}$  as  the equilibrium flow.
  \end{lma}
  
 The proof of Lemma \ref{l:1} uses techniques of Non Zero Sum Games (NZSG) \cite{BsrOlsd}.  We solve the game by explicitly computing the best responses and finding their intersection (refer \cite{uni_hal}). 
\begin{prop}
\label{p:PoA}
The  \textbf{Price of Anarchy} of two atomic users on LB network (Fig. \ref{fig:lb}) with the associated elbow latency function of \eqref{e:Tl} and with identical demands is 
given by
\bear\label{e:PoA}
PoA=\left[1+ \frac{(\frac{  r}{2}-\zeta_m)c_m}{  r L}\right].
\eear
The PoA can be arbitrarily large for large enough $m$.
\end{prop}

The proof is based on Lemma \ref{l:1} (see \cite{uni_hal} for details). 
\vspace{-2mm}
\subsection{Selfish Routing Game with Non-Atomic Users}
\vspace{-1mm}
We consider large number of selfish users at each nodes. The demand of each user is assumed very small such that choice of a single user has negligible impact on the system performance. The latency observed by a user is the sum of the link latency that a user traverse to reach to the destination node. Each user would choose a route so as to minimize its latency at equilibrium. 
\begin{prop}
\label{l:2}
The {\em Price of Anarchy} for the routing game on the LB network with non-atomic selfish users and associated latency given by \eqref{e:Tl} and with identical demands is equal to $1$ for any $m$. 
\end{prop}
The bounded PoA is achieved because the user's use only their local links. This is because the latency observed by a user via a cross link is larger as compared to the case when he chose his local link. However, such direct intuition is not possible in case of atomic users. The proof of Lemma \ref{l:2} is based on variational inequality at Wardrop equilibrium in the similar lines of \cite{rou_bad} (see \cite{uni_hal}).  

Interestingly, the social welfare for selfish routing games with atomic users is unbounded whereas in similar settings it corresponds to social optimum for non-atomic users.  
\vspace{-2mm}
\section{Altruistic Routing Game}
\label{s:alt}
In this section, we consider routing games when users are not entirely selfish, but rather altruistic. Each user minimizes the perceived cost, a weighted sum of his own cost and other users' cost (we proposed this model in \cite{altruism_wiopt}). The weight coefficient is called \emph{degree of Cooperation} (DoC).

\begin{definition}\label{def:coop}
Let $\alpha^i_k$ be the \emph{degree of cooperation} of user $i$ with user $k$.  The perceived cost is a convex combination of the users' cost from the set $\cal I$ as follows:
\vspace{-2mm}
 \bear \label{e:Jhat}
\hat{J}_i(\mathbf{x},\vec \alpha_i)= \sum_{k\in {\cal I}}
\alpha^i_k J_k(\mathbf{x}) ; \,\, \sum_k \alpha^i_k =1
,i=1,...|{\cal I}|.\eear
\end{definition}
 We can express the DoC vector of user $i$  as $\overrightarrow{\alpha ^i}= (\alpha ^i_1,\cdots,\alpha ^i_{|{\cal
I}|})$. Based on the DoC, we can view the
following properties for user $i$,
\begin{itemize}
\item Non-cooperative: if $\alpha^i_k=0,\,\, \textrm{for all}\,k\neq i$. Denote the DoC of a selfish user by $\vec e_i=\{0,\cdots,e_i^i,\cdots,0 \}$, where $e_i^i=1$.
\item Altruistic: User $i$ is fully cooperative with all users and does not care for his benefits, i.e., $\alpha^{i}_i=0 $.
\item Equally-cooperative: if $\alpha^i_j=\frac{1}{|{\cal{I}}|},\, \forall j\in \cal I$, user $i$
is equally cooperative with each user.
\end{itemize}

Note that, although a user cooperates with others, he minimizes his own perceived cost. Therefore, this formulation allows us to analyze the game using the conventional techniques of non cooperative game theory, e.g., Nash equilibrium, etc.
\vspace{-1mm}
\subsection{Unilateral Altruism}
\vspace{-1mm}
We call user $i$ \textbf{unilaterally altruistic},  if he cooperates with other users irrespective of whether the other users are cooperative or  non cooperative with user $i$.  
More precisely, let all users $j\neq i$ be selfish and user $i$ be unilaterally altruistic with degree of cooperation  vector $\vec\alpha_i$, then $\vec \alpha=\{ \vec e_1,\cdots, \vec\alpha_i,\cdots,\vec e_{|\cal I |}\}$. 
The perceived  cost by user $i$ and user $j\neq i$ can be  expressed as
\vspace{-2mm}
\bear \label{e:per_cost}
\hspace{-10mm}&&\hat{J}_i(\mathbf{x},\vec \alpha_i)=\sum_{k \in \cal I} \alpha_k^i J_k (\mathbf{x}),\quad
\hat{J}_{j}(\mathbf{x},\vec e_j)={J}_j(\mathbf{x}).
\vspace{-3mm}
\eear

Let $\textbf{x}^{SE}$ denote the set of Nash equilibrium flow vectors for the selfish routing game. Then, the best selfish Nash equilibrium flow can be given by $\mathbf{\underline x}^{SE}=\argmin_{\textbf{x}\in\textbf{x}^{SE}}{\hat J}_i(\mathbf{ x},\vec e_i)$. Similarly, let $\mathbf{x}^{AL}_{\vec \alpha_i}$ be the set of Nash equilibrium flows\footnote{Existence of equilibrium with unilateral altruism is guaranteed since the strategy set is compact and continuous, and the cost function is concave \cite{BsrOlsd}. } (using \eqref{e:per_cost}) when user $i$ is unilaterally altruistic with DoC $\vec \alpha_i$ and all the other users are selfish. Recall that, an altruistic player truly commits his strategy to be altruistic such that the other players can change their strategy in accordance to this belief. 

\begin{definition}[\textbf{VoU}]
\label{d:vou}
  We quantify the benefit of unilateral altruism by the term {\bf Value of Unilateral Altruism (VoU)}, which is defined as the ratio of the best selfish equilibrium cost  to his best equilibrium cost  when he is unilaterally altruistic. Mathematically, we can  express
\vspace{-2mm}
\bear \label{e:vou}
{VoU}(i)= \frac{{J}_i(\mathbf{\underline x}^{SE})}{\inf_{(\vec \alpha_i,\mathbf{x}^{AL}_{\vec \alpha_i}) }{J}_i(\mathbf{x}^{AL}_{\vec \alpha_i})}
\eear 
\end{definition}
Notice that the equilibrium is obtained based on minimization of perceived cost but the $VoU$ is related to the actual cost incurred by the user. 
By VoU we identify the bound on the improvement that a user may achieve by becoming altruistic (if altruism is beneficial for the user).
Note that, when there exists unique equilibrium with an altruistic user, the ${VoU}$ corresponds to the $\vec \alpha_i$ that minimizes the equilibrium cost. However, when there exist multiple equilibria, the ${VoU}$ corresponds to the best equilibria at its corresponding $\vec \alpha_i$.
By definition, we have $\mathbf{\underline x}^{SE}\subseteq \mathbf{x}^{AL}_{\vec \alpha_i}$, therefore the feasible values of $VoU(i)$ must be $\geq1$. The case of ${VoU}(i)=1$ indicates, either no improvement or degradation with altruism as compared to the selfish situation. The more interesting is the case of $VoU(i)>1$, which corresponds to the situation when a unilaterally altruistic user improves his cost as compared to he being selfish. This is non-intuitive, as  an altruistic user is generally expected to incur more cost.   In the following, we illustrate that one can achieve drastic improvement by being unilaterally altruistic.

\begin{prop}
\label{p:uni}
Suppose user $1$ is altruistic in a two user routing game on a LB Network (Fig. \ref{fig:lb}) with the associated latency given by \eqref{e:Tl} and with identical demand. The ``Value of unilateral altruism"  of user $1$,  $ {VoU}(1)$, is arbitrarily large for sufficiently large $m$.
\end{prop}

Below we provide a brief sketch of the proof (refer \cite{uni_hal} for details).
The selfish users incur unbounded cost at equilibrium (unique),  see Lemma \ref{l:1}. 
 Therefore, a bounded equilibrium cost of user $1$ with his unilateral altruism implies the proposition. However, there can be multiple equilibria with altruism but due to unbounded $J_1^{SE}(.)$, the $VoU(1)$ still remains unbounded even if there exist other better equilibrium. Using these arguments we show $VoU(1)$ is unbounded. In fact, we show that 
$(x_1^{1^*},x_2^{2^*})=(r,r-\delta)$ is an equilibrium with unilateral altruism for a set of $\alpha\in \Gamma$. The users cost is unbounded for this equilibrium, which concludes the proof.

Note that by showing unilateral altruism, not only the altruistic user is able to improve his cost by keeping it bounded, rather the other selfish user also enjoy his improved cost which also remains bounded. This way, even the selfish user also get drastic benefit due to altruistic user's action.

\vspace{-1mm}
\section{VoU for $N$ players}
\label{s:est}
From the discussion of previous sections, one might think that the benefit of altruism (the unbounded VoU) is restricted to two user scenario and it may disappear for a larger number of users. In this spirit we investigate for $n$ user game in this section. We are able to construct a LB type network of  $n+1$ nodes, shown in Figs. \ref{fig:lb}, on which the property of unbounded $VoU$ is retained when a user is unilaterally altruistic. 

Consider that $n$ users (a user at each node) have identical demands to be dispatched  to a common destination node (marked as $n+1^{th}$ node). We compute the selfish routing game performance and subsequently study the performance with an altruistic user. 
 It is intuitive that if all users turn altruistic then the efficiency of selfish equilibrium can be improved. In the earlier sections we studied the game with two users. However, it is more complex to study for $n$ users. We chose a symmetric network, which is more tractable. We find that even in case of $n$ users, the benefit of a single user's altruism is significant. We formally establish our findings with the help from the following lemma and proposition.
 
\begin{figure}[htbp]
\begin{center}
 \resizebox{5cm}{3cm}{\input{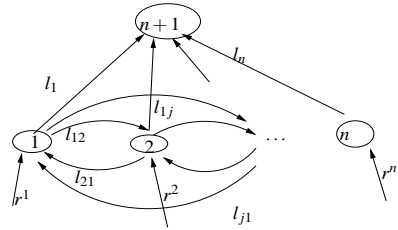}}
 \vspace{-2mm}
 \caption{A LB type network  with $n$ source nodes and a destination node.}
 \label{fig:n_lb}
\end{center}
\vspace{-5mm}
\end{figure}

\begin{lma}
\label{l:n_sel}
The PoA for the $n$ player selfish routing game on a LB network (Fig. \ref{fig:n_lb}) with the latency function as in \eqref{e:Tl}, and with identical demands is arbitrarily large for sufficiently large $m$.
\end{lma}
The proof follows by first showing unique equilibrium by transforming the network into a two node network with parallel links. Exploiting the symmetric structure, we obtain the NEP for selfish users and also for the social optimum. 

\begin{prop} \label{p:n_al}
Consider a user (say user $1$) being unilaterally altruistic while other $n$ users are selfish in a LB network (Fig. \ref{fig:n_lb}), where all other users are selfish users with identical demands and latency as in \eqref{e:Tl}.  The ``Value of unilateral altruism'' of the altruistic user, $VoU(1)$, is arbitrarily large for sufficiently large $m$.
\end{prop}
The proof follows the similar techniques as that in proposition \ref{p:uni} and using lemma \ref{l:n_sel}. 
\vspace{-2mm}
\section{Concluding Remarks and Perspective}
\label{s:conclusion}

Our investigation in this work reveals the benefits of unilateral altruism in a network routing game. We showed with the help of  a simple example with two users that if one user becomes  altruistic, his equilibrium utility drastically improves. An improvement by unilateral altruism is itself non-intuitive. Moreover, the improvement by altruism is drastic, which is quite surprising. To quantify this we proposed a metric, ``Value of Unilateral Altruism". Latter, we also show that the benefit of unilateral altruism is visible for a larger number of players.
 
\vspace{-2mm}
\bibliographystyle{IEEEtran}
\bibliography{routingpara}
{
\newpage
\appendix
{

\subsection{Proof for Lemma \ref{l:1}}
  \Prf
  The Nash equilibrium for the selfish game corresponds to the solution to the following minimization problem 
  \bear
  {\cal P}_2 :\quad 
\textbf{x}^{i^*}=\argmin_{\textbf{x}^i\in\textbf{X}^i} J_i(\textbf{x}^i,\textbf{x}^{-i}), \,\forall\,\, i=1,2.
  \eear
 The cost of user $1$ and $2$ in the above equation can be expressed as follows:
  \bear
&&  J_1(\mathbf{x})=x_1^1 T_1(x_1) +\bar x_1^1(T_{12}(\bar x^1_1)+T_2(x_2)), \\
&& J_2(\mathbf{x})=x_2^2 T_2(x_2) +\bar x_2^2(T_{21}(\bar x^2_2)+T_1(x_1)).
  \eear
  where $\bar x_i^i=r-x_i^i$ for $i=1,2$.
  
The compactness of the strategy set $\mathbf{X}$ with the concavity and continuity of the objective function $J_i(.)$ imply the existence of a Nash equilibrium (see \cite[Thm. 4.4]{BsrOlsd}). A formal proof of existence is provided in \cite{altruism_wiopt}. We provide the proof of unique Nash equilibrium for a general $n$ node LB network ($n$ user) in Lemma \ref{l:n_sel}. Therefore, we conclude that there exist unique equilibrium for the problem ${\cal P}_1$, as it is a special case of Lemma \ref{l:n_sel} with $n=2$. Therefore, if we can compute a Nash equilibrium, then that is the only Nash equilibrium. We can observe that a symmetric Nash equilibrium exist due the symmetric structure of ${\cal P}_1$. In the following, we compute the symmetric Nash equilibrium. 

\subsubsection{Computation of symmetric Nash equilibrium}
We compute the best response for each player to compute the equilibrium. The best responses can be given by the  following expressions
\bear
\label{e:BR}
\begin{aligned}
BR_1(x_2^2)=\argmin_{0\leq x^{1^*}\leq r_1} {J}_1(x^1,x^2)=\max\{\min\{0, a^*\},  r\}, &\\
BR_2(x_1^1)=\argmin_{0\leq x^{2^*}\leq r_2} {J}_2(x^1,x^2)=\max\{\min\{0, b^*\},  r\}.&
\end{aligned}
\eear
where (with some abuse of notation\footnote{$a^*$ is represented in short for $a^*(x_2^2)$ and $b^*$ in short for $b^*(x_1^1)$.}) we denote the \textit{internal solutions} by $a^*$ and $b^*$. By the \textit{internal solution}, we mean  $0<a^*(\textrm{or } b^*)<r$ when   $a^*\Rightarrow \frac{\partial J_1(\mathbf{x})}{x_1^1}=0$ and $b^*\Rightarrow \frac{\partial J_2(\mathbf{x})}{x_2^2}=0$. Equating the above and doing some simple calculus we obtain,
\bear \label{e:br}
\begin{aligned}
a^*(x_2^2)&=\frac{F(x_2)+(c_m+  r g(x_2))}{G(x)},\\
b^*(x^1_1)&=\frac{F(x_1)+(c_m+  r g(x_1))}{G(x)}
\end{aligned}
\eear
where,  $F(x_2)=T_{2}(x_2)-T_{1}(x_1)=-F(x_1)$, $g_1(x_1)=\frac{\partial T_{l_1}(x_1)}{\partial x_1}$, $g_2(x_2)=\frac{\partial T_{2}(x_2)}{\partial x_2}$, and $G(x)=g_1(x_1)+g_2(x_2)$. The gradient of $T_1(x)$ at $  r$ is $g_1( r)=L/  \delta_{m}$. 

Exploiting the symmetry  we note $x_1=x_2=  r,\,g(x_1)=g(x_2)=g(  r)$, and 
$ F(x_1)=F(x_2)=0$ at equilibrium.  Therefore, the equilibrium flow can be given by
 \bear
 \label{eq:X11*}
 x_1^{1^*}=x_2^{2^*}=\frac{c_m}{2g(  r)}+\frac{  r g(  r)}{2g(  r)}=\frac{  r }{2}+\zeta_m
 \eear
 where $\zeta_m=\frac{1}{2}\left(\frac{c_m}{L/  \delta_m}\right)$.  
 The Nash equilibrium flow for selfish users is expressed as $\textbf{x}^{SE}=\{\frac{  r }{2}+\zeta_m,\frac{  r }{2}+\zeta_m \}$. Note that $\zeta_m\geq0$ for large enough $m$.

2) Using \eqref{eq:X11*}, the equilibrium cost can be given by
\bear \nonumber
 &&\hspace{-10mm}J_1^{SE}=\left(\frac{  r}{2}+\zeta_m\right)T_1(x_1)+\left(\frac{  r}{2}-\zeta_m\right)\left(T_{12}\left(\frac{  r}{2}-\zeta_m\right)\right.\\
&& \hspace{-2mm} +T_2(x_2)\Big)=r L + \left(\frac{  r}{2}-\zeta_m\right) c_m.
\eear
This concludes the proof.
\endpf
}

\subsection{Proof for Proposition \ref{p:PoA}}
  \Prf
Recall that  the definition of  price of anarchy from \eqref{eq:PoA}
\vspace{-2mm}\bear\vspace{-2mm}
PoA=\sup_{\textbf{x}\in \textbf{x}^{SE}} \frac{\sum_i J_i(\textbf{x})}{\sum_i J_i(\textbf{x}^{OPT})} \quad.
\eear

One can easily observe that at social optimum (OPT) only local links are used. Therefore, the user cost $J^{OPT}_i$ at social optima is given as \vspace{-2mm}
\bear \label{e:OPT}
J_1^{OPT}=  r T_1(  r)=rL.
\eear
Therefore, the price of anarchy for any $m$ can be given by 
\bear
PoA=\frac{r L + \left(\frac{  r}{2}-\zeta_m\right) c_m}{rL}=\left[1+ \frac{(\frac{  r}{2}-\zeta_m)c_m}{  r L}\right].
\eear
Note that $c_m$ is an increasing sequence and $\zeta_m$ is a decreasing sequence. Hence, $PoA$ can be arbitrary large for large enough $m$. 
\endpf

\subsection{Proof of Lemma \ref{l:2}}
\Prf
Consider non-atomic type users at source node $1$ and source node $2$ with a total demand rate of $r$ each to be dispatched to the destination node $3$. 

Let $p\in \cal P$ denote a path that a user can take from a source node to destination node and $\textbf{T}_p(.)$ denote the cumulative latency observed for the path. The concept of road traffic wardrop equilibrium, originally introduced by Wardrop\cite{wardrop}, which is the non-cooperative equilibrium for non-atomic users in network traffic, is defined in terms of variational inequality as 
\bear \label{e:wardrop}
 \textbf{T}_p(\textbf{x})-A\geq0; \,\, ({\textbf{T}}_p(\textbf{x})-A)x^i_p=0,\,\, \for \,\,p\in \cal P,
\eear
where $A=\min_{p\in \cal P} \textbf{T}_p(\textbf{x})$.

The Wardrop equilibrium falls into the category of potential games with an infinite number of users and can be expressed by a single convex optimization\cite{altman:a}. Hence, there exist a unique equilibrium.  

There are two possible paths from source node $1$ to destination node: a) through $l_1$; and, b) through $l_3\rightarrow l_2$.  Path  Similarly, from source node $2$ following are the feasible paths: c) through $l_2$; and, d) through $l_4\rightarrow l_1$. 
Using \eqref{e:wardrop}, we can easily see that at equilibrium the set ${\cal P}=\{a,c\}$, i.e., only local links are used.
Therefore, the equilibrium flow is given by $\textbf{x}^{*}=\{x_{1^*},x_{2^*} \}=\{r,r\}$ and the latency cost experienced by a user is $T_1(.)=L$ and total cost of users from a source node equals $\sum_{k=1}^r T_1(.)=rL$. 

Clearly, the user's cost is bounded and equal to the social optimum at Wardrop equilibrium. 
%
%
%
\endpf

\subsection{Proof of Proposition \ref{p:uni}}
{
\Prf In an altruistic routing game, users minimize their perceived cost ($\hat J_i(.)$). It is interesting to note that although the  perceived cost is minimized non-cooperatively, the effect of altruism appears due the coupling through the cost. Therefore, the Nash equilibrium of this two user game can be expressed by the following equivalent problem: 
\bear \label{e:pmin}
{\cal P}_2 :\quad 
x^i=\argmin_{\textbf{x}\in\textbf{X}} \hat J_i(\textbf{x}), \,\forall\,\, i=1,2.
\eear
where the perceived cost of users (when user $1$ is altruistic) can be expressed as
\bear
\begin{aligned}
\hat{J}_1(\textbf{x})&=  (1-\beta)J_1(\textbf{x})+ \beta J_2(\textbf{x}),\\
\hat{J}_2(\textbf{x})&= J_2(\textbf{x}).
\end{aligned}
\eear
 Existence of an equilibrium is direct from the compact strategy set $\textbf{X}$ and concave and continuity of the cost function $\hat J_i(\textbf{x})$ \cite{BsrOlsd}.
 
 In this altruistic equilibrium, we are interested in an equilibrium which can improve the user's cost as the selfish equilibrium cost is very large. It turns out that the cost becomes unbounded or large only when a significant amount of flow is routed thorough cross link, which is the main reason of inefficiency. Therefore, a bounded cost can be achieved only when one can restrict almost all the flow to one's local link. In this spirit an altruistic game can achieve the equilibrium flow to be $(x_1^1,x_2^2)=(r,r-\delta)$ for at least some $\beta$, because it can result in bounded cost. 

We call $x_i^i=r$ as ``load taker'' strategy because user $i$ do not reciprocate (sends back flow) to any flow sent to him, i.e., takes all the load. Similarly, $x_i^i=r-\delta$ is the so called ``delta load taker" strategy as it only pushes enough flow to the other link so that his local link gets to the knee, provided the other player doesn't push back.  We aim to show that $(x_1^1,x_2^2)=(\textrm{load taker}, \textrm{delta load taker})=(r,r-\delta)$ is an
 equilibrium strategy.  we use the best response approach to show the equilibrium for large enough $m$. We user $\delta$ and $\delta_m$ alternatively in the proof for ease of notation.
 

 
\textbf{Computation of $BR_2(x_1^1)$} :
The cost of $u_2$ for a load taker strategy of user $1$ is given by (let denote $p=x_2^2$ and $\bar p=r-x_2^2$),
\bear \label{e:J2p}
{J}_2(p)
=\begin{cases}
 \bar{p}[T_{12}(\bar{p})+T_1(r+\bar{p})] & \textrm{if }{p} \leq r-\delta,\\
 pT_2(p)+\bar{p}[T_{12}(\bar{p})+T_1(r+\bar{p})] & \textrm{if }{p} >r-\delta.
\end{cases}
\eear 
Notice that the ${J}_2(p)$ has discontinuity only at $x_2^2=r-\delta$. The marginal cost can be given as 
\bear \label{e:231}
\frac{\partial {J}_2(p)}{\partial p}=
\begin{cases}
A(p)<0 & \textrm{if } p \leq r-\delta,\\
{B}_m(p)>0 & \textrm{if } p >r-\delta,
\end{cases}
\eear
Using \eqref{e:J2p}, we have $A(p)=-\bar p\frac{L}{\delta^m}-[d_m+\frac{L}{\delta_m} (\bar p)+L]<0,$ and $B_m(p)= \frac{(2p-r)L}{\delta_m}-d_m>(r-2\delta_m)\frac{L}{\delta_m}-d_m$.
%
%
%
We note that $B_m(p)>0$ for sufficiently large $m$.   Hence,  $BR_2(r) =r-\delta$.

\textbf{Computation of $BR_1(x_2^2)$}:  The cost of $u_1$ for a delta load taker strategy of $u_2$ is given by (let denote $x_1^1=p$ and $x_2^2=\bar\delta=r-\delta$),
\bear \nonumber
\hat J_1(p)&=&(1-\beta)J_1(\textbf{x})+\beta J_2(\textbf{x})\\
\nonumber&=&(1-\beta)[pT_1(p+\delta)+\bar p(T_{12}(\bar p)+T_2(\bar p + \bar \delta))]\\
&&+\beta[\bar \delta T_2(\bar p+\bar \delta)+\delta (T_{21}(\delta)+T_1(p+\delta))]
\label{e:J1p}
\eear
Noting that $\hat J_1(p)$ is strictly convex in $p\in[0,r]$, if $\frac{\partial \hat J_1}{\partial p}<0$ at $p=r$ then $\frac{\partial \hat J_1}{\partial p}<0$ for $p\in[0,r]$.  Therefore, we examine if $\frac{\partial \hat J_1}{\partial p}<0$ holds true at $p=r$. Taking the derivative of $\hat J_1(p)$, we have 
\bear \label{e:J1d} \nonumber
&&\hspace{-9mm}\frac{\partial \hat J_1(p)}{\partial p}=(1-\beta)[p(g_1(p+\delta)+g_2(\bar p+\bar \delta))+T_1(p+\delta)-T_2(\bar p+\bar \delta)\\
  &&\hspace{-5mm}-(rg_2(\bar p+\bar \delta)+T_{12}(\bar p))]+\beta[\delta g_1(p+\delta))-(\bar \delta) g_2(\bar p+\bar \delta) ] 
\eear
At $ p =r^{-}$ (less than and close to $r$), we have 
$g_1(p+\delta)=g_2(\bar \delta+\bar p)=\frac{L}{\delta_m}$, $T_1(p+\delta)=L+(p+\delta-r)\frac{L}{\delta_m}$, and $T_2(\bar \delta+\bar p)=L+(r-\delta-p)\frac{L}{\delta_m}$. Plugging the values in \eqref{e:J1d}, we obtain 
\bear
\frac{\partial \hat J_2(p)}{\partial p}<0, \, \for \beta\in \Gamma
\eear
where the set $\Gamma$ turns out to be a non empty set, given by $(\frac{1}{2} ,1]$.
Therefore, we can state that $BR_1(r-\delta)=r, \for \, \beta\in\Gamma$.

Therefore, together it constitute the Nash equilibrium. Denoting the equilibrium flow by $\tilde{ \textbf{x}}^{AL}$, and cost of user $1$ as $J_1(\tilde{ \textbf{x}}^{AL})$, we have
\vspace{-2mm}
 \bear
\tilde{ \textbf{x}}^{AL}&=&(r,r-\delta),\,\, \for\,\, \beta\in\Gamma, \\
J_1(\tilde{ \textbf{x}}^{AL})&=&rT_1(r+\delta)=2rL . \label{e:J1AL}
\eear 
The set of equilibria is denoted by $\textbf{x}^{AL}$ when there exist multiple equilibria. Denote $\textbf{\underline x}^{AL}\in \textbf{x}^{AL}$ such that 
\bear \label{e:xalt} J_1(\textbf{\underline x}^{AL})<J_1(\mathbf{x}), \textrm{for any} \, \mathbf{x} \in \textbf{x}^{AL}. \eear

\textbf{Computation of $VoU$}: Using \eqref{eq:X11*} and \eqref{e:J1AL}, and the definition \ref{d:vou}, we can express  
\bear
\nonumber {VoU}(1)&=& \frac{{J}_i(\mathbf{\underline x}^{SE})}{\inf_{(\vec \alpha_i,\mathbf{x}^{AL}_{\vec \alpha_i}) }{J}_i(\mathbf{x}^{AL}_{\vec \alpha_i})}>\frac{J_1(\textbf{x}^{SE})}{J_1(\tilde{\textbf{x}}^{AL})}\\
&&=\frac{r L+( r/2-\zeta_m)c_m}{2rL}. \label{e:vou1}
\eear  
The last inequality is from \eqref{e:xalt}. The quantity $\frac{r L+( r/2-\zeta_m)c_m}{2rL}$ can be arbitrarily large for large $m$.
%
%
This concludes the proof.

\endpf
}

{
\subsection{Proof of Lemma \ref{l:n_sel} }
\label{proof:n_sel}
\Prf 
In order to compute PoA, we need to compute the worst equilibrium cost for selfish user and the cost at social optimum. To do so, 
We  go in the following steps: 
\begin{enumerate}
\item[1.] We show that there exist unique Nash equilibrium for selfish users. 
\item[2.] Since there can be only one equilibrium, we exploit the symmetric property and compute the Nash equilibrium. Obviously, this can be used for worst equilibrium cost of PoA computation. 
\item[3.] Finally, we note that the social optimum is unique, the PoA follows by directly comparing the ration of selfish equilibrium and social optimum. 
\end{enumerate}
In the following, we go step by step.

We allowed arbitrary number of node and a user associated with each user, as shown in Figs. \ref{fig:n_lb}. The Nash equilibrium for the selfish game corresponds to the solution to the following minimization problem 
  \bear
  {\cal P}_1 :\quad 
x^{i^*}=\argmin_{\textbf{x}\in\textbf{X}} J_i(\textbf{x}), \,\forall\,\, i=1,\cdots,n.
  \eear
 The cost of user a $i$ in the above equation can be expressed as:
  \bear \nonumber
 J_i(\mathbf{x})&=&x_i^i T_i(x_i) +\bar x_i^i(T_{ij}(\bar x^i_i)+T_j(x_j)),\\
 &&\hspace{5mm}\for \,i\neq j=1,\cdots,n.\hspace{1mm}
\eear
Where, $\bar x_i^i=r^i-x_i^i$ .

We show the existence of unique Nash equilibrium by transforming our problem into parallel link network problem with arbitrary number of users, for which it known to exist unique Nash equilibrium (see \cite[Theorem 1.]{orda}).

Consider a network consisting of two nodes: $a$ and $b$, and of $I$ parallel directed links, all from node $a$ to node $b$. There are $i$ users, $i=1,\cdots,I$, having node $a$ as the source node and node $b$ as the destination node. The total demand rate for user $i$ is $r^i$. 

Let $x_l^i$ be the rate at which user $i$ sends over the link $l,\, (l=1,\cdots,I$, which satisfy the positive flow constraint, $x_l^i\geq 0, \,\, \forall  l$ and $\sum_l x_l^i=r^i$. User $i$ determines $(x_1^i,\cdots,x_I^i)$ so as to minimize his cost $\overline{ J}_i(\overline{ \textbf{x}})$ where $\overline{ \textbf{x}}=(x_l^i, l\in(1,\cdots,I), i\in {\cal I})$. Define $x_l=\sum_i x_l^i$.

We assume that $\overline{ J}(\overline{ \textbf{x}})$ is the sum of the link cost functions: $\overline{ J_i}(\overline{ \textbf{x}})= \sum_l \overline{ J_i^l}(\overline{ \textbf{x}})$, where $\overline{ J_i^p}(\overline{ \textbf{x}})$ are expressed in terms of the cost $J_i^l(\textbf{x})$, defined in the original problem, as follows. For any $l$ and any $i$,
\bears
\overline{ J}_{i}^i(\overline{ \textbf{x}})&=&J_i^i(x_i^i,x_i)\\
\overline{ J_i^j}(\overline{ \textbf{x}})&=& J_i^{ij}(x^i_j,x_j)+J^j_i(x_j^i,x_j),\,\, \forall \, j\neq i.
\eears 
recall that $J_i^i$ represent cost of user $i$ in its local link $i$ and the corresponding flow is $x_i^i$.   


We note that the cost function of the original problem satisfies the following assumptions (as that in \cite{orda}):
\begin{enumerate}
\item [G1.] $J_i(\textbf{x})$ is the sum of the link costs incurred in each link, i.e.,
\vspace{-3mm}
\bears
J_i(\textbf{x})=\sum_l J_i^l(\textbf{x})
\eears 
\item[G2.] $J_i^l(\textbf{x})$ is continuous (piecewise continuous with our elbow latency function, which suffice the condition) functions whose range is the nonnegative quadrant and their image is $[0,\infty]$.
\item[G3.] $J_i^l$ are convex functions in the rate sent by user $i$ over link $l$.
\item[G4.] Whenever finite, $J_i^l$ is continuously differentiable in the flow sent by user $i$ to link $l$. The marginal cost is denoted by $K_i^l(\textbf{x})=\frac{\partial J_i^l(\textbf{x})}{\partial x_i^l}$.
\item[G5.] If not all users have finite cost and one user user has infinite cost then it can change its own flow o make this cost finite. This ensures that at any equilibrium users have finite cost if it is feasible. 
\end{enumerate}
The following assumption is also satisfied by $J_i(\textbf{x})$, which is very important for uniqueness (\cite{orda}):  
\begin{enumerate}
\item[$\Pi_1$] $K_i^l(.)$ is function of two arguments: i) the total flow on the link $l$, $x_l$ and ii) the flow of user $i$ on link $l$, $x_i^l$. $K_i^l(x^l_i,x_i)$ is strictly increasing in both of its arguments. 
\end{enumerate}

\subsubsection{{Uniqueness of Nash equilibrium}}
\label{s:unique}
If the cost of original problem with LB network satisfy the assumption $(\Pi_1)$, it follows that the costs for the new routing problem also do satisfy the assumption $(\Pi_1)$. The new routing problem has a unique Nash equilibrium under the assumptions $(\Pi_1)$ (see Theorem 1 in \cite{orda}). By identifying the decision variable $\overline{\textbf{x}}$ in the new routing problem with the decision variable $\textbf{x}$ of the original problem with LB network, we see that the minimization problem faced by each user is same in both the cases, and therefore we conclude that user optimum in our problem is also unique. Note that the although we considered a symmetric network, the proof holds good for non symmetric LB network also.

\subsubsection{Computation of  equilibrium}
Since all the users have identical demands and facing identical network conditions, the users will have identical strategy at equilibrium. Such equilibrium are often called as symmetric equilibrium. 
Let $x_i^i$ be the equilibrium strategy of user $i$, the flow in his local link, and $x_i^j$ is the flow in a cross link to user $j$. 

Let us compute the best response of user $i$ for the strategy of user $j\neq i$. Let $x_i^i$ be user $i$'s best response for the strategy of user $j$ be $x_j^j$. Note that we consider that all the other user's have identical strategy($x_j^j$).   Therefore, the best response of user $i$ is given by
\bear
BR_i(x_j^j,n)=\argmin_{\textbf{x}_i\in \textbf{X}} J_i(\textbf{x}).
\eear
The cost of user $i$ is given by,
\bears
&&\hspace{-8mm}J_i(x_i^i,x_j^j)= x_i[T_i(x_i^j + (r-x_j^j))]\\
&&\hspace{-6mm}+ \sum_{j\neq i} \frac{\bar x_i^i}{n-1}[c_m(\frac{\bar x_i^i}{n-1})+ T_j(x_j^j +\frac{(r-x_i^i)}{n-1})+\frac{(n-2)(r-x_k^k)}{n-1})].
\eears
where $k\neq i,j$. The derivative of $J_i(.)$ is (denote $p=x_i^i$)
\bears
&&\hspace{-10mm}\frac{\partial J_i(p,x_j^j)}{\partial p}= p g_i(p+r-x_j^j)+T_i(p+r-x_j^j)\\
&&\hspace{-9mm}-\bar p\left[g_j\left(x_j^j+\frac{(r-p)}{n-1}+\frac{r-x_k^k}{n-1}\right)\right]- \left[c_m+T_j\left(x_j^j +\frac{(n-2)\bar x_k^k+ \bar p}{n-1}\right)\right],\\
&&\hspace{-9mm}=p[g_i(x_i)+g_j(x_j)]+[T_i(x_i)-T_j(x_j)]-(r g_j(x_j)+c_m).
\eears
Due to symmetric structure, we note that $x_i=x_{(j\neq i)}=r$. Thus, we have
$T_i(x_i)-T_j(x_j)=0$ and $g_i(x_i)=g_j(x_j)=g$. Therefore, 
\bears
BR_i(x_j^j)=\frac{c_m+r g}{2g}=\frac{1}{2}\left[r+\frac{c_m}{L/\delta_m}\right]=r/2+\zeta_{m}.
\eears
Since all the users are symmetric, the equilibrium strategy of any user is same. Therefore,
the equilibrium strategy and the equilibrium cost of a user $i$ can be given by
\bear
\begin{aligned}
x_i^{i^{SE}}&=r/2+\zeta_{m},\\
J_i^{SE}&=r L + \left(\frac{  r}{2}-\zeta_m\right) \left(c_m\right).
\end{aligned}
\eear

One can directly observe that at social optimum (OPT) only local links are used. Therefore, the user cost $J^{OPT}_i$ at social optima is given as
\bear
J_i^{OPT}=  r T_i(r)=rL,\,\forall \,\,i.
\eear

\subsubsection{Computation of  PoA}
Therefore the price of anarchy can be expressed as
\bear
PoA=\sup_m \frac{\sum_i r L + (\frac{  r}{2}-\zeta_m) \left(c_m\right) }{\sum_i r L}=1+\sup_m \frac{(\frac{r}{2}-\zeta_m) \left(c_m\right)}{ r L}.
\eear 
Evidently, the PoA can be arbitrarily large for sufficiently large $m$.
 This concludes the proof.
\endpf

}

{
\subsection{Proof of Proposition \ref{p:n_al} }
\Prf The equivalent problem for the $n$ user altruistic routing game considered can be expressed as (similar as in proposition \ref{p:uni}) \vspace{-3mm}
\bear \label{e:pnmin}
{\cal P}_n :\quad 
x^i=\argmin_{\textbf{x}\in\textbf{X}} \hat J_i(\textbf{x}), \,\forall\,\, i=1,\cdots,n.
\eear
where the perceived cost of users (when user $i$ is altruistic) can be given by 
\vspace{-2mm}
\bear
\hat{J}_i(\textbf{x})&=  \sum_{(k=1)}^n \alpha_k^i J_k(\textbf{x}).
\eear
 Existence of an equilibrium is direct from the compact strategy set $\textbf{X}$ and concave and continuity of the cost function $\hat J_i(\textbf{x})$ \cite{BsrOlsd}. 
 Following the approach of proposition \ref{p:uni}, we show that the load taker strategy of altruistic user and delta load taker strategy of other users constitutes a Nash equilibrium. Since this equilibrium yields a bounded cost for altruistic user, $VoU$ is be arbitrarily large for large $m$.
 
 Let  index altruistic user by $i$ and selfish users by $j(\neq i)=1,\cdots,n$. We compute the best response for both type users.
 
\textbf{Computation of $BR_j(x_i^i=r)$} :
The cost of a selfish user ($u_j$) for a load taker strategy of user $i$ is given by (let denote $p=x_j^i$, $\bar p=r-x_j^i=x_j^j$, and $x_j^k=q$, $k\neq i,j$),
\bears
J_j(\bar p)&=& \bar{p}[T_j(\bar p)+(n-2)q T_k(\bar p)+(n-2)qT_{jk}(q)\\
&&+pT_i(r+(n-1)p)+p T_{ji}(p)]
\eears
Since the latency function is discontinuous, the cost function can be re-expressed as follows
\bear \label{e:Jjp}
\hspace{-2mm}{J}_j(\bar p)
\hspace{-1mm}=\hspace{-1mm}\begin{cases}\hspace{-1mm}
p[T_i(r+(n-1) p)+c_m]+(n-2)q c_m &\textrm{if }{\bar p} \geq r-\delta,\\
 \bar p T_j(\bar p)+(n-2)q T_k(\bar p)+(n-2)qc_m&\\
 +p[T_i(r+(n-1)p)+T_{ji}(p)] &\textrm{if }{p} >r-\delta.
\end{cases}\hspace{-6mm}
\eear 
Noting the discontinuity only at $x_j^j=r-\delta$, the marginal cost can be given as 
\bear \label{e:2311}
\frac{\partial {J}_j(\bar p)}{\partial \bar p}=
\begin{cases}
A^1(\bar p)<0 & \textrm{if }\bar p \geq r-\delta,\\
{B}^1_m(\bar p)>0 & \textrm{if }\bar p <r-\delta,
\end{cases}
\eear
Using \eqref{e:Jjp}, we have $A^1(\bar p)=- p\frac{L}{\delta^m}-pT_i(r+(n-1)p)<0,$ and $B^1_m(\bar p)= \frac{(2\bar p-r)L}{\delta_m}+(n-2)\frac{L}{\delta_m}-c_m-(T_i(r+(n-1)p)-T_j(\bar p))$.
We note that $B^1_m(p)>0$ for sufficiently large $m$.   Hence, the \textbf{best response is a delta load taker} strategy, i.e., $BR_j(r) =r-\delta$.

\textbf{Computation of $BR_i(x_j^j=r-\delta)$}:  The cost of the altruistic user ($u_i$) for a delta load taker strategy of $u_j$ is given by (let denote $x_i^i=p$ and $x_j^j=r-\delta=\bar\delta$), \vspace{-2mm}
\bear \nonumber
&&\hspace{-9mm}\hat J_i(p)=(1-\beta)J_i(\textbf{x})+(n-1)\frac{\beta}{(n-1)} J_j(\textbf{x})\\
\nonumber&&\hspace{-7mm}=(1-\beta)[pT_i(p+(n-1)\delta)+(n-1)\frac{\bar p}{n-1}(T_{ij}(\frac{\bar p}{n-1})\\
\nonumber&&\hspace{-6mm}+T_j(\frac{\bar p}{n-1} + \bar \delta))]+\beta[\bar \delta T_{j}(\frac{\bar p}{n-1}+\bar \delta)\\
&&\hspace{24mm}+\delta (T_{ji}(\delta)+T_i(p+\delta(n-1)))].
\label{e:Jip}
\eear
Noting that $\hat J_i(p)$ is  convex in $p\in[0,r]$, if $\frac{\partial \hat J_i}{\partial p}<0$ at $p=r^-$ (less than and close to $r$) then $\frac{\partial \hat J_i}{\partial p}<0$ for $p\in[0,r]$.  Therefore, below we examine if $\frac{\partial \hat J_i}{\partial p}<0$ holds true at $p=r^-$. The derivative of $\hat J_i(p)$ is  
\bear \label{e:Jid} \nonumber
&&\hspace{-9mm}\frac{\partial \hat J_i(p)}{\partial p}=(1-\beta)[p(g_i(p+(n-1)\delta)+g_j(\frac{\bar p}{n-1}+\bar \delta))\\
\nonumber&&\hspace{-6mm}+T_i(p+(n-1)\delta)-T_j(\frac{\bar p}{n-1}+\bar \delta)-(rg_j(\frac{\bar p}{n-1}+\bar \delta)\\
  &&\hspace{-6mm}+T_{ij}(\frac{\bar p}{n-1}))]+\beta[\delta g_i(p+\delta(n-1)))-(\bar \delta) g_j(\frac{\bar p}{n-1}+\bar \delta)].\hspace{5mm} 
\eear
%
Plugging the values of $g(.)$ and $T(.)$ in \eqref{e:Jid}, we obtain 
\bear
\frac{\partial \hat J_i(p)}{\partial p}<0, \, \for \beta\in \Gamma
\eear
where the set $\Gamma=(\frac{1}{n} ,1]$. 
Hence, we can state that $BR_i(r-\delta)=r, \for \, \beta\in\Gamma$.

Therefore, together it constitute the Nash equilibrium. Denoting the equilibrium flow by $\tilde{ \textbf{x}}^{AL}$, and cost of user $i$ as $J_i(\tilde{ \textbf{x}}^{AL})$, we have
\vspace{-2mm}
 \bear
\tilde{ \textbf{x}}^{AL}&=&(r,r-\delta),\,\, \for\,\, \beta\in\Gamma, \\
J_i(\tilde{ \textbf{x}}^{AL})&=&rT_i(r+\delta)=rL . \label{e:JiAL}
\eear 
The set of equilibria is denoted by $\textbf{x}^{AL}$ when there exist multiple equilibria. Denote $\textbf{\underline x}^{AL}\in \textbf{x}^{AL}$ such that 
\bear \label{e:xalt} J_i(\textbf{\underline x}^{AL})<J_i(\mathbf{x}), \textrm{for any} \, \mathbf{x} \in \textbf{x}^{AL}. \eear

\textbf{Computation of $VoU$}: Using \eqref{eq:X11*} and \eqref{e:JiAL}, and the definition \ref{d:vou}, we can express  
\bear
\nonumber {VoU}(i)&=& \frac{{J}_i(\mathbf{\underline x}^{SE})}{\inf_{(\vec \alpha_i,\mathbf{x}^{AL}_{\vec \alpha_i}) }{J}_i(\mathbf{x}^{AL}_{\vec \alpha_i})}>\frac{J_1(\textbf{x}^{SE})}{J_i(\tilde{\textbf{x}}^{AL})}\\
&&=\frac{r L+( r/2-\zeta_m)c_m}{2rL}. \label{e:vou1}
\eear  
The last inequality is from \eqref{e:xalt}. The quantity $\frac{r L+( r/2-\zeta_m)c_m}{2rL}$ can be arbitrarily large for large $m$.
This concludes the proof.

\endpf

}

}
\end{document}